\documentclass[prd,showpacs,preprintnumbers,amsmath,amssymb]{revtex4}

\usepackage{graphicx}
\usepackage{dcolumn}
\usepackage{bm}

\def\be{\begin{eqnarray}}
\def\ee{\end{eqnarray}}
\def\bea{\begin{eqnarray}}
\def\eea{\end{eqnarray}}

\def\0T{{\bf 0}_\perp}
\begin{document}


\title{Transverse Force on Quarks in DIS}

\author{Matthias Burkardt}
 \affiliation{Department of Physics, New Mexico State University,
Las Cruces, NM 88003-0001, U.S.A.}

\date{\today}

\begin{abstract}
The interaction dependent part of the $x^2$-moment of the twist-3 polarized 
parton distribution $g_2(x)$ is related to the transverse
force acting on the active quark in deep-inelastic scattering
off a transversely polarized nucleon
immediately after it has absorbed the virtual photon.
Similarly, the $x^2$-moment of the chirally odd twist-3 unpolarized 
parton distribution $e(x)$ can be related to the transverse
force experienced by a transversely polarized quark 
ejected from a transversely polarized nucleon.
\end{abstract}

\maketitle
\section{Introduction}
In the analysis of high-energy scattering processes,
leading twist effects are often both easier to isolate (e.g. by
increasing $Q^2$ until higher twist effects are negligible) 
and have a 
more direct physical interpretation than contributions from higher 
twist operators which are often intermingled with $\frac{1}{Q^n}$ 
corrections to leading twist operators. A notable exception
is the twist-3 polarized structure function 
$g_2(x,Q^2)$, which can be cleanly separated from its twist-2 
counterpart $g_1(x,Q^2)$ when scattering longitudinally polarized
leptons from a transversely polarized target. While in 
general, the contribution from $g_2$ to the polarized cross section
is suppressed by powers of $\frac{1}{{Q^2}}$, at a target polarization
angle of $90\,^\circ$ the leading
contribution vanishes and the contribution from $g_2$ is exposed
without kinematical suppression. This property of the polarized
DIS cross section thus allows a clean extraction of 
higher twist matrix elements, without the need for fitting and
subtracting a leading twist effect, and thus makes
polarized deep-inelastic scattering (DIS) a rare opportunity for
studying higher twist effects (for an overview, see Ref. 
\cite{Jaffe:Erice}).

As $g_2(x,Q^2)$ involves higher twist, it does not have a parton
interpretation as a single particle density. Indeed, the twist-3 
part of $g_2$ is related to quark-gluon correlations whose
intuitive interpretation may not be immediately clear. Since 
$g_2(x,Q^2)$ is related to (electromagnetic) 
polarizabilities at low $Q^2$, these twist-3 matrix elements have
been called color polarizabilities in the literature \cite{Ji}.
However, at high $Q^2$, the twist-3 piece of $g_2(x,Q^2)$ is 
described by a local correlator and the physical interpretation
as a polarizability no longer applies. 
Indeed, while nucleons need to be
polarized in order to study $g_2(x,Q^2)$, the nucleons are not 
distorted, but only `spin-alligned'. The quark-gluon correlations
embodied in the twist-3 part of $g_2(x,Q^2)$ are then obtained as
a matrix elements of a certain operator in a spin-alligned, but
undeformed, nucleon. This is very different from the usual use of the
term `polarizability' as the tendency of a charge or magnetization
distribution to be distorted from its normal shape by an external
field. Of course, one could broaden the notion of `polarizability'
to encompass matrix elements that are only non-zero when
the nucleon is polarized, but within such a broadened definition,
other spin-dependent observables, such as the polarized
parton distribution $\Delta q(x)$ or even the magnetic moment
of the nucleon, would then also become `polarizabilities' in the
broader sense. 

The main purpose of this paper is to explore an alternative physical 
interpretation of these particular twist-3 matrix elements as a {\em force}. 
First we summarize the connection between the $x^2$ moment
of $g_2(x,Q^2)$ and quark-gluon correlations. After discussing the
connection between these correlations and the transverse force on
the active quark in DIS, we then estimate sign and magnitude of that 
force based on DIS data, lattice calculations and heuristic pictures.

\section{$x^2$ moments and quark-gluon correlations}
The chirally-even spin-dependent twist-3 parton distribution 
$g_2(x)=g_T(x)-g_1(x)=\frac{1}{2}\sum_q e_q^2 g_2^q(x)$ is defined as\begin{eqnarray}
\int \frac{d\lambda}{2\pi}e^{i\lambda x}
\langle PS|\bar{\psi}(0)\gamma^\mu\gamma_5\psi(\lambda n)
|_{Q^2}|PS\rangle
=2\left[g_1(x,Q^2)p^\mu (S\cdot n) 
+ g_T(x,Q^2)S_\perp^\mu +M^2 g_3(x,Q^2)n^\mu (S\cdot n)
\right].
\end{eqnarray}
where $p^\mu$ and $n^\mu$ are light-like vectors along the $-$ and
$+$ light-cone direction with $p\cdot n=1$. 

Using the equations
of motion $g_2^q(x)$ can be expressed as a sum of a piece that is
entirely determined in terms of $g_1^q(x)$ plus an interaction
dependent twist-3 part that involves quark gluon correlations
\cite{WW}
\be
g_2^q(x)&=&g_2^{q,WW}(x)+\bar{g}_2^q(x) \label{eq:WW}\\
g_2^{q,WW}(x)&=&-g_1^q(x)+\int_x^1 \frac{dy}{y}g_1^q(y).\nonumber
\ee
Here we have neglected $m_q$ for simplicity. For example,
the $x^2$ moment yields
\be
\int dx x^2 \bar{g}_2^q(x)= \frac{d_2^q}{3} \label{eq:d2}
\ee
with  \cite{Shuryak,Jaffe}
\be
2 M {P^+}P^+ S^x d_2^q
=g\left\langle P,S \left|\bar{q}(0)\gamma^+G^{+y}(0)q(0) 
\right|P,S\right\rangle
\label{eq:twist3}.
\ee
$d_2^q$ is related to the experimentally measured polarized structure functions
through
\be
d_2\equiv \frac{1}{2} \sum_q e_q^2 d_2^q = 3\int dx \bar{g}_2(x)x^2
= \int dx\left[2g_1(x)+3g_2(x)\right] x^2.
\ee

In the limit where $Q^2$ is so low that the virtual photon 
wavelength is larger
than the nucleon size,  the electro-magnetic
field associated with the two virtual photons appearing
in the forward Compton amplitude corresponding to the structure 
function is nearly homogenous accross the nucleon and the
spin-dependent structure function $g_2(x,Q^2)$ can be related to
spin-dependent polarizabilities.
In contradistinction, in the Bjorken limit, the matrix elements
describing the moments of $g_2(x,Q^2)$ are given by local
correlation functions, such as (\ref{eq:twist3}).
Nevertheless, because of the abovementioned low $Q^2$ interpretation
of $g_2$, the {\em local} matrix elements appearing in 
(\ref{eq:twist3})  
\be
\chi_E 2M^2 {\vec S} = \left\langle P,S\right|
q^\dagger {\vec \alpha} \times g {\vec E} q \left| P,S\right\rangle
\quad\quad\quad\quad\quad\quad\quad\quad
\chi_B 2M^2 {\vec S} = \left\langle P,S\right|
q^\dagger g {\vec B} q \left| P,S\right\rangle,
\label{eq:chi}
\ee
where
\be
d_2 = \frac{1}{4}\left(\chi_E+2\chi_B\right),
\label{eq:d2chi}
\ee
(note that $\sqrt{2}G^{+y}=B^x-E^y$)
are sometimes called color electric and 
magnetic polarizabilities \cite{Ji}. In the following we will 
discuss why, at high $Q^2$, a better interpretation for these 
matrix elements is that of a `force'.

In electro-magnetism, the $\hat{y}$-component of the Lorentz 
force $F^y$ acting on a particle with unit charge moving, with 
(nearly) the speed of light along the $-\hat{z}$ direction,
${\vec v}\approx (0,0,-1)$, reads
\be
F^y = \left[ {\vec E} + {\vec v}\times {\vec B}\right]^y
= \left(E^y - B^x\right) = -\sqrt{2}G^{+y},
\ee
which involves the same linear combination of
Lorentz components that also appears in the 
gluon field strength tensor in (\ref{eq:twist3}).
This simple observation already suggests a connection between
$d_2$ and the color Lorentz force on a quark that moves
(in a DIS experiment) with ${\vec v}\approx (0,0,-1)$.
We therefore propose a new semi-classical interpretation for the matrix
element appearing in the definition of $d_2$ (\ref{eq:twist3}) as
the average transverse force acting on the struck quark in DIS in the instant after it has been hit by the virtual photon.

In order to explore this connection further we compare the
matrix element defining $d_2$ with that describing the
average transverse momentum of quarks in semi-inclusive DIS
(SIDIS) \cite{sivers}.
The average intrinsic transverse momentum of quarks bound in a
nucleon vanishes and therefore any net transverse momentum
of quarks in a SIDIS experiment must come
from the final state interactions (FSI) \cite{collins}.
The average transverse
momentum of the ejected quark (also averaged over the momentum
fraction $x$ carried by the active quark in order to render the
matrix element local in the position of the quark field operator)
in a SIDIS experiment can thus be represented by the matrix element
\cite{QS,JiYuan,Pijlman}
\be
\langle k_\perp^y\rangle =  -\frac{1}{\sqrt{2}M}
\left\langle P,S \left|\bar{q}(0)\gamma^+\int_0^\infty dx^-gG^{+y}(x^+=0,x^-)
q(0) \right|P,S\right\rangle,
\label{eq:QS}
\ee
where Wilson-line gauge links along $x^-$
are implicitly understood, but not
written out explicitly. 

The matrix element appearing in (\ref{eq:QS}) thus has a simple 
physical interpretation as the transverse impulse obtained
by intergrating the color Lorentz force along the trajectory of 
the active quark
--- which is an almost light-like trajectory along the 
$-\hat{z}$ direction, with $z=-t$.

Since $z=-t$ for a particle moving along the light-cone, one finds
$dt=\frac{dx^-}{\sqrt{2}}$.
In order to make the correspondence more explicit,
we now rewrite Eq. (\ref{eq:QS}) as an integral over time
\be
\langle k_\perp^y\rangle = - \frac{\sqrt{2}}{2P^+}
\left\langle P,S \right|\bar{q}(0)\gamma^+\int_0^\infty \!\!\!dt\, gG^{+y}(t,z=-t)
q(0) \left|P,S\right\rangle
\label{eq:QS2}
\ee
in which the physical interpretation of $- \frac{\sqrt{2}}{2P^+}
\left\langle P,S \right|\bar{q}(0)\gamma^+G^{+y}(t,z=-t)
q(0) \left|P,S\right\rangle$ as being the (ensemble
averaged) transverse force 
acting on the struck quark at time $t$ after being struck by the
virtual photon becomes more apparent. 
In particular,
\be
\label{eq:QS3}
F^y(0)&\equiv& - \frac{\sqrt{2}}{2P^+}
\left\langle P,S \right|\bar{q}(0) \gamma^+G^{+y}(0)
q(0) \left|P,S\right\rangle\\
&=& -{\sqrt{2}} MP^+S^xd_2
= -{M^2}d_2,
\nonumber
\ee
where the last equality holds only in the rest frame 
($P^+=\frac{1}{\sqrt{2}}M$) and for $S^x=1$,
can be interpreted as the averaged transverse
force acting on the active quark
in the instant right after it has been struck by the virtual photon.

A measurement of the $x^2$-moment $f_2$ of the 
twist-4 distribution $g_3(x)$ 
allows determination of the expectation value of a different
linear combination of 
Lorentz/Dirac components of the quark-gluon correlator appearing
in (\ref{eq:twist3}) \cite{f2}
\be
f_2 M^2S^\mu = \frac{1}{2} \left\langle p,S\right|
\bar{q}g\tilde{G}^{\mu \nu}\gamma_\nu q\left|p,S\right\rangle . 
\ee
Using rotational invariance, to relate various Lorentz components
one thus finds a linear combination of the matrix elements of
electric and magnetic quark-gluon correlators (\ref{eq:chi})
\be
f_2=\chi_E-\chi_B,
\ee
that differs from that in (\ref{eq:d2chi}).
In combination with (\ref{eq:twist3}) this 
allows a decomposition of the force into electric and magnetic
components $F^y= F^y_E+F^y_B$, using
\be
F_E^y(0)= - \frac{M^2}{4} \chi_E\quad \quad \quad \quad
F_B^y(0)= - \frac{M^2}{2} \chi_B
\ee
for a target nucleon polarized in the $+\hat{x}$ direction, where
\cite{Ji,color}
\be
\chi_E = \frac{2}{3}\left(2d_2+f_2\right)
\quad \quad \quad \quad
\chi_B = \frac{1}{3}\left(4d_2-f_2\right).
\ee

\section{Chirally-odd twist-3 matrix elements}
A relation similar to (\ref{eq:QS3}) can be derived for the
$x^2$ moment of the twist-3 scalar PDF $e(x)$. For its
interaction dependent twist-3 part $\bar{e}(x)$ one finds for an
unpolarized target \cite{Yuji} 
\be
4MP^+P^+ e_2 &=& -\sum_{i=1}^2
g\left\langle p\right|\bar{q}\sigma^{+i}G^{+i}q
\left|P\right\rangle,
\label{eq:odd1}
\ee
where $e_2\equiv \int_0^1 dx x^2\bar{e}(x)$.

Semiclassically, the matrix element on the r.h.s. of Eq. (\ref{eq:odd1})
can be related to the average transverse force acting on
a transversely polarized quark in an unpolarized target right after 
being struck by the virtual photon.
Indeed, the matrix elements of $\frac{1}{2}\bar{q}\left(\gamma^+-\sigma^{+y}\right)q$
yield the density of quarks polarized in the $+\hat{x}$ direction. Since the
matrix element of $\bar{q}\left(\gamma^+G^{+y}\right)q$ vanishes for an unpolarized
target, we can thus relate Eq. (\ref{eq:odd1}) to the average color Lorentz force
(right after absorbing the virtual photon)
in the $\hat{y}$ direction for quarks polarized in the $+\hat{x}$ direction as
\be
F^y(0) = \frac{1}{2\sqrt{2}P^+} g\left\langle P\right| 
\bar{q} \sigma^{+y}G^{+y}q\left|
P\right\rangle = -\frac{1}{\sqrt{2}}MP^+S^x e_2 =- \frac{M^2}{2} e_2,
\label{eq:Fe2}
\ee
where the last identity holds only in the rest frame of the target
nucleon and for $S^x=1$. In the physical interpretation of 
(\ref{eq:Fe2}) it is important to keep in mind that, for a given 
flavor, the number of quarks on which the force in (\ref{eq:Fe2}) acts is 
only half that in (\ref{eq:QS3}) as only half the quarks in an
unpolarized nucleon will be polarized in the $+\hat{x}$ direction and we thus normalized only by a factor $P^+$ instead of $2P^+$ as in (\ref{eq:QS3}).

Since the quark-gluon correlator in (\ref{eq:Fe2}) is local, the force interpretation
applies to the moment immediately after the virtual photon is absorbed.

The above interpretation is also consistent with matrix elements describing transverse
single-spin asymmetries in a SIDIS experiment. For transversely polarized quarks in
an unpolarized target, this asymmetry is described by the Boer-Mulders function
$h_1^\perp(x,{\bf k}_\perp^2)$ \cite{BM,trento}.
For the
average transverse momentum in the $+\hat{y}$ direction,
for a quark polarized in the $+\hat{x}$ direction
(${\bf k}_\perp^2$ moment of the
Boer-Mulders function integrated also over $x$), reads
\be
\langle k^y \rangle = \int dx \int d^2{\bf k}_\perp {\bf k}_\perp^2 
h_1^\perp(x,{\bf k}_\perp^2)=
\frac{1}{2P^+}\int_0^\infty dx^-
g\left\langle p\right| \bar{q}(0) \sigma^{+y}G^{+y}(x^-)q(0)\left|
p\right\rangle
\label{eq:odd2}.
\ee
Since the $x^-$ integral represents the trajectory of the ejected quark,
the interpretation of the first integration point in (\ref{eq:odd2}) with the force immediately after absorbing the virtual photon is consistent with (\ref{eq:Fe2}).

Inspired by the above observations one might be tempted to anticipate that a similar result also exists for the $x^2$ moment of the interaction dependent twist-3 part of 
$h_L(x)$. However, the $x^2$ moment of $h_L$ contains only a contribution from
$h_1$  and there is no interaction dependent piece. 

\section{Discussion}
When the target nucleon is transversely polarized, e.g. in the 
$+\hat{x}$ direction the axial symmetry in the transverse plane
is broken. In particular, the quark distribution (more precisely
the distribution of the $\gamma^+$-density that dominates in DIS
in the Bjorken limit) in the transverse plane is deformed 
\cite{IJMPA}. The average deformations can be related to the
contribution from each quark flavor to the anomalous magnetic moment
of the nucleon and was predicted to be quite substantial
\cite{IJMPA} and has also been observed in lattice QCD
\cite{hagler}.

Given the fact that, for a nucleon polarized in the $+\hat{x}$
direction the $\gamma^+$-distribution for $u$ ($d$) is
shifted towards the $\pm\hat{y}$ direction suggests that
these quarks also `feel' a nonzero color-electric force pointing
on average in the $\mp\hat{y}$ direction, i.e. one would
expect that $d_2$ is positive (negative) for $u$ ($d$) quarks.
This is also consistent with a negative (positive) sign for the 
proton Sivers 
function as observed by the HERMES collaboration \cite{hermes}
and the vanishing Sivers function for deuterium in the
COMPASS experiment \cite{compass}.
In fact, in the large $N_C$ limit, one would expect that $d_2$
for $u$ ($d$) quarks are equal and opposite. Note that while
$d_2$ for $u$ ($d$) quarks being exactly equal and opposite would
imply the same for protons (neutrons), any deviation from
being exactly equal and opposite is enhanced for proton (neutron)
since there is a significant cancellation between the two quark 
flavors in the nucleon. 

If all spectators in the FSI were to `pull' in the same direction, 
the force on the active quark would be of the order of the QCD
string tension $\sigma \approx (450MeV)^2$, which would translate
into a value $d_2\sim 0.2$. However, it is more natural to expect
a significant cancellation between forces from
spectators pulling the active quark in different directions, the
actual value of $d_2$ is probably much
smaller, i.e. $0.1>d_2> 0.01$ appears to be more natural.
Instanton based models have suggested an even smaller value
\cite{Weiss}.

Heuristic arguments/lattice calculations \cite{mb:brian,hagler} 
also suggest that the deformation of 
(the $\gamma^+$-distribution for) transversely polarized
quarks in an unpolarized nucleon is more significant than that of
unpolarized quarks in a transversely polarized nucleon. When
applied to the final state interactions, this observation suggests
that the transverse force for transversely polarized quarks in an unpolarized nucleon
is stronger than that for unpolarized quarks in a transversely polarized nucleon.
$|e_2|>|d_2|$ (the fact that in an unpolarized nucleon only half 
the quarks are polarized in the $+\hat{x}$-direction is compensated
by the factor $\frac{1}{2}$ in (\ref{eq:Fe2}).


Lattice calculations of the twist-3 matrix element yield 
\cite{latticed2}
\be
d_2^{(u)} = 0.020 \pm 0.024 \quad \quad \quad \quad
d_2^{(d)} = -0.011 \pm 0.010
\ee
renormalized at a scale of $Q^2=5$ GeV$^2$ for the smallest
lattice spacing in Ref. \cite{latticed2}.
Note that we have multiplied the numerical results from 
\cite{latticed2} by a factor of $2$ to account for the different
convention for $d_2$ being used.

Here the identity $M^2\approx 5$GeV/fm is useful
to better visualize the magnitude of the force.
\be
F_{(u)} = -100 \pm 120 {\rm MeV/fm}\quad \quad \quad \quad
F_{(d)} = 56 \pm 52 {\rm MeV/fm}.
\ee
This result is consistent with the chromodynamic lensing picture \cite{lensing}, which suggests
that $F_{(u)}$ and $F_{(d)}$ are of about the same order of magnitude and with
opposite sign. The same holds in the large $N_C$ limit.
A vanishing Sivers effect for an isoscalar target would be more
consistent with equal and opposite average forces. However, since
the error bars for $d_2$ include only statistical errors, the 
lattice result may not even be inconsistent with 
$d_2^{(d)} \sim - d_2^{(u)}$. Up to date lattice calculations for $d_2$ are long overdue.

The average transverse momentum from the Sivers effect is
obtained by integrating the transverse force to infinity
(along a light-like trajectory) 
$\langle k^y\rangle = \int_0^\infty dt F^y(t)$
(\ref{eq:QS2}). This motivates us to
define an `effective range' 
\be
R_{eff} \equiv \frac{\langle k^y\rangle}{F^y(0)}.
\ee
Note that $R_{eff}$ depends on how rapidly the correlations fall
off along a light-like direction and it may thus be larger than
the (spacelike) radius of a hadron. 
Although strictly speaking, unless the functional form of the integrand is known, 
$R_{eff}$ cannot really tell us about the range of the FSI. However if the integrand does not oscillate, $R_{eff}$ may still provide a rough
estimate about the actual range of the FSI.

From fits of the Sivers function 
to SIDIS data
\cite{Mauro} one finds approximately $|\langle k^y\rangle|\sim
100$ MeV \cite{Mauro}. 
Together with the (average) value for $|d_2|$ from
the lattice this translates into an effective range $R_{eff}$ of more than
one fm.
It would be interesting to compare $R_{eff}$ for different quark 
flavors and as a function of $Q^2$, but this requires more
precise values for $d_2$ as well as the Sivers function.

Note that a complementary approach to the effective range was 
chosen in Ref. \cite{Stein}, where the twist-3 matrix element
appearing in Eq. (\ref{eq:QS3}) was, due to the lack of lattice 
QCD results, estimated using QCD sum rule techniques. Moreover,
the `range' was taken as a model {\em input} parameter to estimate
the magnitude of the Sivers function.

The impact parameter distribution for quarks polarized in
the $+\hat{x}$ direction was found to be shifted in the
$+\hat{y}$ direction \cite{DH,hagler,engel,mb:brian}.
Applying the chromodynamic lensing model implies a force
in the negative $-\hat{y}$ direction for these quarks and one
thus expects $e_2<0$ for both $u$ and $d$ quarks. 

It would be interesting to study not only whether the effective range
is flavor dependent, but also whether there is a difference between 
the chirally even and odd cases. It would also be very interesting
to learn more about the time dependence of the FSI by calculating
matrix elements of $\bar{q}\gamma^+ \left(D^+ G^{+\perp}
\right)q$, or even higher derivatives, in lattice QCD.
Knowledge of not only the value of the integrand at the origin,
but also its slope and curvature at that point, would be very
useful for estimating the integral in Eq. (\ref{eq:QS}).

Although the force interpretation for the $x^2$ moments of twist-3 PDFs
is strictly speaking only valid semi-classically, it is nevertheless important
to also understand the scale dependence of this interpretation. The PDFs and hence also
their $2^{nd}$ moment depend on the QCD scale (the latter
decreases with $Q^2$). On the other hand 
one would also expect that the force acting on the escaping quark depends on the scale:
at low $Q^2$, with few sea quarks or additional gluons present, the nucleon wave function still has long range color coherence. In contradistinction, at high $Q^2$, with
additional $\bar{q}q$ pairs and gluons present, the nucleon wave function will
exhibit color coherence only over a very short range. In particular, the
'chromodynamic lensing' mechanism \cite{lensing} should be suppressed at high $Q^2$. 

\section{Summary}

The quark-gluon correlations embodied in the $x^2$-moment $d_2$ of the interaction
dependent 
twist-3 part of the polarized PDF $g_2$ can be identified with the transverse component 
of the color-Lorentz force acting on the struck quark in the instant 
after absorbing the virtual photon. The direction of the the
force for $u$ and $d$ quarks can be understood in terms of the
transverse deformation of parton distributions for a transversely
polarized target. In combination with a measurement of the $x^2$ 
moment of the twist-4 polarized PDF $g_3$ one can even decompose 
this force into color-electric and magnetic components.
Although still quite uncertain, first eperimental/lattice results
suggest values on the order of $50-100$MeV/fm for the net force.
This should be compared with the net transverse momentum due to
the Sivers effect which is on the order of $100$MeV 
\cite{Mauro}.

The $x^2$ moment $e_2$ of the chirally odd twist-3 (scalar) PDF $e(x)$
can be related to the transverse force acting on transversely
polarized quarks in an unpolarized target. Therefore,
$-e_2$ is to the Boer-Mulders function $h_{1}^\perp$, what
is $d_2$ to the Sivers function $f_{1T}^\perp$.

{\bf Acknowledgements:}
I would like to thank 
A.Bacchetta, D. Boer, J.P. Chen, Y.Koike, and Z.-E. Meziani for 
useful discussions. 
This work was supported by the DOE under grant numbers 
DE-FG03-95ER40965 and DE-AC05-06OR23177 (under which Jefferson
Science Associates, LLC, operates Jefferson Lab).

\end{document}